\font\BBig=cmr10 scaled\magstep2
\font\BBBig=cmr10 scaled\magstep3


\def\title{
{\bf\BBBig
\centerline{Vanishing of the conformal anomaly}\bigskip
\centerline{for strings in}\bigskip
\centerline{a gravitational wave}
}}

\def\foot#1{
\footnote{($^{\the\foo}$)}{#1}\advance\foo by 1
} 
\def\ccr{\cr\noalign{\medskip}}


\def\authors{
\centerline{C. DUVAL\foot{D\'epartement de Physique, Universit\'e
d'Aix-Marseille II and
Centre de Physique 
\hfill\break
Th\'eorique
 CNRS-Luminy, Case 907, F-13288 MARSEILLE,
Cedex 09 (France)
\hfill\break
e-mail:duval@cpt.univ-mrs.fr.}
Z. HORV\'ATH\foot{Institute for Theoretical Physics,
E\"otv\"os University, H-1088 BUDAPEST,
\hfill\break
 Puskin u. 5-7 (Hungary)
e-mail: zalanh@ludens.elte.hu}
P. A. HORV\'ATHY\foot{D\'epartement de Math\'ematiques,
Universit\'e de Tours, Parc de Grandmont,
\hfill\break
F-37200 TOURS (France) e-mail: horvathy@univ-tours.fr}}
}

\def\runningauthors{Duval, Horv\'ath, Horv\'athy}

\def\runningtitle{Vanishing anomaly \dots}


\voffset = 1cm 
\baselineskip = 14pt 

\headline ={
\ifnum\pageno=1\hfill
\else\ifodd\pageno\hfil\tenit\runningtitle\hfil\tenrm\folio
\else\tenrm\folio\hfil\tenit\runningauthors\hfil
\fi
\fi}

\nopagenumbers
\footline = {\hfil} 


\font\tenb=cmmib10 
\newfam\bsfam

\textfont\bsfam=\tenb
\mathchardef\betab="080C
\mathchardef\xib="0818
\mathchardef\omegab="0821
\mathchardef\deltab="080E
\mathchardef\epsilonb="080F
\mathchardef\pib="0819
\mathchardef\sigmab="081B
\mathchardef\bfalpha="080B
\mathchardef\bfbeta="080C
\mathchardef\bfgamma="080D
\mathchardef\bfomega="0821
\mathchardef\zetab="0810


\def\and{\qquad\hbox{and}\qquad}
\def\where{\qquad \hbox{where} \qquad}
\def\kikezd{\parag\underbar}

\def\smallcirc{{\raise 0.5pt \hbox{$\scriptstyle\circ$}}} 
\def\smallover#1/#2{\hbox{$\textstyle{#1\over#2}$}} %
\def\2{{\smallover 1/2}}
\def\ccr{\cr\noalign{\medskip}}
\def\parag{\hfil\break} 

\def\={\!=\!}
\def\-{\!-\!}

\def\G{{\Gamma^+}}

\newcount\ch 
\newcount\eq 
\newcount\foo 
\newcount\ref 

\def\chapter#1{
\parag\eq = 1\advance\ch by 1{\bf\the\ch.\enskip#1}
}

\def\equation{
\leqno(\the\ch.\the\eq)\global\advance\eq by 1
}

\def\reference{
\parag [\number\ref]\ \advance\ref by 1
}

\ch = 0 
\foo = 1 
\ref = 1 


\title
\vskip .5cm
\authors
\vskip .5cm

\parag
{\bf Abstract.}\hskip 2mm

{\it Using the non-symmetric-connection approach proposed by Osborn, we
demonstrate that, for a bosonic string in a specially chosen
plane-fronted gravitational wave and an axion background, the conformal
anomaly vanishes at the two-loop level. Under some conditions,
 the anomaly vanishes at all orders.}

\vskip.5cm
\noindent{April 1993.
 Physics Letters {\bf B 313}, 10 (1993).}

\chapter{Introduction.}

In a recent paper [1] Osborn carried out a two-loop calculation of the
$\beta$-functions of a bosonic string in the presence of a non-trivial
background metric $g_{\mu\nu}$ and of an antisymmetric axion field
$b_{\mu\nu}$. He used dimensional regularization, extending the
two-dimensional antisymmetric tensor
$\epsilon_{\sigma\tau}$ into a complex structure in $2n$ dimensions. An
immediate consequence of his procedure was to combine the metric and the
axion into a single, non-symmetric tensor $t_{\mu\nu}$; the three-form
$H_{\mu\nu\rho}$ associated to the axion appears as the {\it torsion} of
the non-symmetric connection.

In this Letter we apply Osborn's conceptual framework to an explicit example,
namely to string propagation in a generalized pp wave (plane-fronted
gravitational wave with parallel rays)
 of the type discussed by Brinkmann [2], whose metric is
$$
g_{\mu\nu}dx^{\mu}dx^{\nu}
=-2dudv+2a_i(u,x)dx^idu+g_{ij}(u,x)dx^idx^j+k(u,x)du^2.
\equation
$$
Brinkmann's metrics (1.1) are more general
than the well-known [3] plane waves in four dimensions, with $a_i\=0$:
a non trivial \lq vector potential' $a_i$ can arise, if the spacetime is
at least five dimensional.
All these metrics admit a covariantly constant null vector,
$\partial/\partial v$,
and, assuming the vacuum Einstein equations $R_{\mu\nu}\=0$
are satisfied, have zero scalar curvature.

String propagation in an ordinary pp wave with flat transverse metric
$g_{ij}\=\delta_{ij}$ and with zero vector potential $a_i\=0$ was
previously considered [4-5], and it
was found that the  vacuum Einstein
equations $R_{\mu\nu}\=0$ imply the vanishing of the conformal anomaly to
 all orders in $\sigma$-model perturbation theory. (Including axions and
dilatons yields  more general conditions [5-7]). In the Appendix of his
conference talk [5] Horowitz shortly discussed the case of a
non-trivial $a_i$, but he concluded that the anomaly does not 
in general vanish.

However, in the above papers, the terms coming from the curvature and from
the axion are treated separately, and are independently set to vanish. The
question arises, therefore, whether some cancellation between the various terms
can take place.
We show here that this can indeed happen: the
vector potential can act as a counterterm, canceling
the contribution to the $\beta$-function
of the axion (at least at the two-loop level).
This is similar to what happens in the Wess-Zumino model [8],
where the conformal invariance at the quantum level can be restored
by adding a topological term with a suitable coefficient.

The calculation is particularly clear in Osborn's unified framework which
 appears hence to be the ideal way to treat
 the higher-order contributions to the $\beta$-function equations.

\goodbreak

\chapter{Non-symmetric connections.}

Let us first outline, following the ideas of Osborn [1], how the
metric and axion fields
can be unified into a single framework,
by allowing for non-symmetric connections. Let
$g_{\mu\nu}$ be a $D$-dimensional Lorentz metric, and $b_{\mu\nu}$
the antisymmetric tensor which corresponds to the axion. Set
$t_{\mu\nu}\=g_{\mu\nu}+b_{\mu\nu}$, and define a
non-symmetric connection by
$$
\G^\rho_{\ \mu\nu}=\Gamma^\rho_{\ \mu\nu}+H^\rho_{\ \mu\nu},
\equation
$$
where the $\Gamma^\rho_{\ \mu\nu}$'s are the Christoffel symbols of the
metric, and
$H\={\smallover1/6}\,H_{\rho\mu\nu}dx^{\rho}\wedge
dx^{\mu}\wedge dx^{\nu}$ is the three-form $\2\,db$,
$H_{\rho\mu\nu}\=\2\left(\partial_\rho b_{\mu\nu}+\partial_\mu
b_{\nu\rho}+\partial_\nu b_{\rho\mu}\right)$.
(Since the non-symmetric quantities $\G$, etc. do not have the
usual symmetry properties, the position of the indices is important).

The metric-Christoffels form the symmetric part,
$\Gamma^\rho_{\ \mu\nu}\=\G^\rho_{\ (\mu\nu)}$,
and the axion field is the
antisymmetric part,
$H^\rho_{\ \mu\nu}\=\G^\rho_{\ [\mu\nu]}$.
Remarkably, the non-symmetric connection is expressed as
$$
\G_{\rho\mu\nu}=\2\left(\partial_\nu t_{\rho\mu}+\partial_\mu
t_{\nu\rho}-\partial_\rho t_{\nu\mu}\right).
\equation
$$

Indices are raised and lowered by the metric,
$g_{\mu\nu}$. The curvature is
$$
{R^+}^\rho_{\ \sigma\mu\nu}=\partial_\mu\G^\rho_{\ \nu\sigma}-
\partial_\nu\G^\rho_{\ \mu\sigma}+\G^{\rho}_{\ \mu\alpha}\G^\alpha_
{\ \nu\sigma}-
\G^{\rho}_{\ \nu\alpha}\G^\alpha_{\ \mu\sigma}.
\equation
$$
$R^+_{\ \ \alpha\beta\gamma\delta}$ is antisymmetric
in the first two as well as in the last two indices.
In terms of the
symmetric and antisymmetric parts,
the non-symmetric Riemann tensor can be also written as
$$
{R^+}^\rho_{\ \sigma\mu\nu}=R^\rho_{\ \sigma\mu\nu}
+\nabla_\mu H^{\rho}_{\ \nu\sigma}-\nabla_\nu H^{\rho}_{\ \mu\sigma}
+H^{\rho}_{\ \mu\alpha}H^{\alpha}_{\ \nu\sigma}
-H^{\rho}_{\ \ \nu\alpha}H^{\alpha}_{\ \ \mu\sigma},
\equation
$$
where $\nabla$ is the metric covariant derivative.
The commutator of the covariant derivatives $\nabla^+$
is now related to the curvature and the torsion. We shall need its
expression for $4$-component tensors
$$\eqalign{
\big[\nabla_\mu^+,\nabla_\nu^+\big]Y_{\alpha\beta\gamma\delta}&=
-2H_{\sigma\mu\nu}{\nabla^+}^{\sigma} Y_{\alpha\beta\gamma\delta}
\cr
&-{R^+}^\sigma_{\ \alpha\mu\nu}Y_{\sigma\beta\gamma\delta}
-{R^+}^\sigma_{\ \beta\mu\nu}Y_{\alpha\sigma\gamma\delta}
-{R^+}^\sigma_{\ \gamma\mu\nu}Y_{\alpha\beta\sigma\delta}
-{R^+}^\sigma_{\ \delta\mu\nu}Y_{\alpha\beta\gamma\sigma},
\cr
}
\equation
$$
showing that $H$ plays indeed the role of a torsion tensor.

As for any connection,
the curvature satisfies a Bianchi identity.
In our case, it can be worked out as
$$
\nabla^+_{[\nu}{R^+}^\alpha_{\ |\lambda|\beta\kappa]}
-2H^\sigma_{[\nu\beta}{R^+}^\alpha_{\ |\lambda|\kappa]\sigma}=0,
\equation
$$
where \lq $|\lambda|$' means \lq no permutation in $\lambda$'.
Since the three-form $H$ is closed, $dH\=0$, the curvature tensor satisfies
the identity
$$
\nabla^+_\mu H_{\nu\rho\sigma}\=\smallover3/2R^+_{[\nu\rho\sigma]\mu},
\equation
$$

The Ricci tensor $R^+_{\mu\nu}\={R^+}^\alpha_{\ \
\mu\alpha\nu}$
is found to be
$$
R^+_{\mu\nu}=R_{\mu\nu}+
\nabla_\alpha H^{\alpha}_{\ \nu\mu}
-H_\mu^{\ \rho\alpha}H_{\nu\rho\alpha}.
\equation
$$

Its symmetric part,
$R_{\mu\nu}-H_\mu^{\ \rho\alpha}H_{\nu\rho\alpha}$,
is recognized here as the
first-order contribution to the metric $\beta$-function, and
the antisymmetric part, $\nabla_\alpha H^{\alpha}_{\ \nu\sigma}$, is the
 contribution for the $b_{\mu\nu}$-$\beta$-function [6, 7].
The first-order anomaly-cancellation condition
is, therefore, unified into
$$
R^+_{\mu\nu}=0,
\equation
$$
which generalizes the standard vacuum Einstein equations to connections with
torsion.

\goodbreak

\chapter{Solvable examples.}

As an example, let us consider the non-symmetric tensor
$$
t_{\mu\nu}=\pmatrix{\delta_{ij}&0&a_i+b_i\ccr
0&0&-2\ccr
a_j-b_j&0&k\cr}
=
\pmatrix{\delta_{ij}&0&a_i\ccr
0&0&-1\ccr
a_j&-1&k\cr}
+\pmatrix{0&0&b_i\ccr
0&0&-1\ccr
-b_j&1&0\cr},
\equation
$$
which clearly describes an axion field
$b_{\mu\nu}(x,u)$, with non-trivial $iu$ components
in a Brinkmann metric
$d{\bf x}^2\!+\!2a_i(x,u)dx^idu\!+\!k(x,u)du^2\-2dudv$.

The components of the connection are readily calculated to give
the Riemann curvature tensor.
The non-zero components of this tensor
are
$$\eqalign{
{R^+}^i_{\ jku}&=\2\partial_k\bigg[{\partial a_l\over\partial x^j}
-{\partial a_j\over\partial x^l}
+{\partial b_l\over\partial x^j}-{\partial b_j\over\partial
x^l}\bigg]\delta^{il},
\ccr
{R^+}^i_{\ ujk}&=-\2\partial_l\bigg[{\partial a_j\over\partial x^k}
-{\partial a_k\over\partial x^j}
-{\partial b_j\over\partial x^k}+{\partial b_k\over\partial x^j}
\bigg]\delta^{il},
\ccr
{R^+}^v_{\ ijk}&=-\2\partial_i\bigg[{\partial a_j\over\partial x^k}
-{\partial a_k\over\partial x^j}
-{\partial b_j\over\partial x^k}+{\partial b_k\over\partial x^j}
\bigg],
\ccr
{R^+}^i_{\ uju}&=\left[\partial_j\Big({\partial a_l\over\partial u}
-\2{\partial k\over\partial x^l}\Big)
-\2\partial_u\bigg({\partial a_l\over\partial x^j}-{\partial
a_j\over\partial x^l}
-{\partial b_l\over\partial x^j}+{\partial b_j\over\partial
x^l}\bigg)\right.
\ccr
&\left.\qquad-\smallover1/4
\bigg({\partial a_l\over\partial x^k}-{\partial a_k\over\partial x^l}
+{\partial b_l\over\partial x^k}-{\partial b_k\over\partial x^l}\bigg)
\bigg({\partial a_k\over\partial x^j}-{\partial a_j\over\partial x^k}
-{\partial b_k\over\partial x^j}+{\partial b_j\over\partial x^k}\bigg)
\right]\delta^{il}.
\cr
}
\equation
$$
as well as those obtained from these by antisymmetry.
The first-order conditions for the anomaly cancellation are, therefore,
$$\left\{\eqalign{
R^+_{ju}&=\2\partial^i\bigg[{\partial a_i\over\partial x^j}
-{\partial a_j\over\partial x^i}
+{\partial b_i\over\partial x^j}
-{\partial b_j\over\partial x^i}\bigg]=0,
\ccr
R^+_{uj}&=-\2\partial^i\bigg[{\partial a_i\over\partial x^j}
-{\partial a_j\over\partial x^i}
-{\partial b_i\over\partial x^j}
+{\partial b_j\over\partial x^i}\bigg]=0,
\ccr
R^+_{uu}&=\partial^i\Big({\partial a_i\over\partial u}-\2{\partial
k\over\partial x^i}\Big)
-\smallover1/4\Big({\partial a^i\over\partial x^k}-{\partial
a_k\over\partial x^i}\Big)^2+\smallover1/4\Big({\partial b_i\over\partial x^k}
-{\partial b_k\over\partial x^i}\Big)^2=0.
\cr}\right.
\equation
$$
(Here the square of $f_{ij}$ means $f_{ij}f^{ij}$).

These equations can be solved. Let, for example, both the vector potential
and the axion field be linear in the transverse
coordinates,
$$
a_i\=\2\,a_{ij}(u)x^j
\and
b_i\=\2\,b_{ij}(u)x^j.
\equation
$$
Then the only non-zero component of $H$ is
$H_{iju}\=\2(b_{ij}\-b_{ji})$. Similarly,
$\partial_ia_j\-\partial_ja_i\=\2(a_{ij}\-a_{ji})$. The two
upper equations in (3.3) are identically satisfied, and the last one
reduces to the Poisson equation
$$
\Delta k=\smallover1/2\big(a_{ij}a^{ij}-b_{ij}b^{ij}\big),
\equation
$$
which generalizes the constraint given in Ref. [5] to the case of a
non-zero vector potential.

So far we have only studied the first-order condition for anomaly cancellation.
Adapting an argument of Horowitz and Steif [5] to our situation,
 we can prove, however,
that, for the linear choice (3.4),
the higher order terms in the perturbation expansion of the anomaly
vanish at {\it all orders}.

These terms are in fact tensors with two indices,
constructed from the Riemann tensor $R^+$, its
covariant derivatives, and from $H$. (The covariant derivatives of
$H$ can be expressed in terms of $R^+$, by Eq. (2.7)). Observe that all
components of the Riemann tensor (3.2) with three spatial
indices are zero because the $a_{ij}$ are now assumed
position-independent. Similarly, $H$ is independent of the transverse
coordinates and its nonvanishing components have an $u$ index.
Therefore, all tensors with two indices, composed only from the
powers of $R^+$ and of $H$, will be zero. The reason is that
these expressions will involve summation for at least one of the
$u$ indices of these tensors, which yields zero, as a
consequence of the particular form of the metric.

As an illustration of
this mechanism, we show here that the \lq square' of the $H$
tensor is zero.
Let us introduce the notation
$H^2_{\mu\nu}\=H_\mu^{\ \rho\kappa}H_{\nu\rho\kappa}$, and let
$H^2\=\big({H^2}\big)_\mu^\mu\=H_{\mu\nu\rho}H^{\mu\nu\rho}$.
The only non-zero component of $\big(H^2\big)_{\mu\nu}$ is
$\big(H^2\big)_{uu}$.
Therefore, $\big({H^2}\big)_u^u\=g^{u\alpha}
\big({H^2}\big)_{u\alpha}\=0$ since $g^{uu}\=0$.
The vanishing of all
higher powers can be shown similarly.

Next, in terms containing $R^+_{\mu\nu\rho\sigma}$ and its derivatives,
one has to contract at least one $u$-index, which can either be an index
of $R^+_{\mu\nu\rho\sigma}$ or of $\nabla^+_{\mu}$. But all terms with
at least one upper $u$ index vanish.
Those terms constructed out of $R^+$ and from $H$ vanish in the same
way.

Finally, ${\nabla^+}^\mu{\nabla^+}^\rho
R^+_{\mu\nu\rho\sigma}$ are related to the covariant derivatives of the
Ricci tensor by the (generalized) Bianchi identity. Contracting Eq.
(2.6) in the indices $\nu$ and $\alpha$, we get in fact
$$
\nabla^+_{\nu}{R^+}^\nu_{\ \lambda\beta\kappa}
-\nabla^+_{\beta}{R^+}_{\lambda\kappa}
+\nabla^+_{\kappa}{R^+}_{\lambda\beta}
+2H_{\beta}^{\ \sigma\nu}R^+_{\nu\lambda\sigma\kappa}
-2H_{\kappa}^{\ \sigma\nu}R^+_{\nu\lambda\sigma\beta}
-2H_{\ \beta\kappa}^{\sigma}R^+_{\lambda\sigma}=0.
\equation
$$
The sum
$2H_{\beta}^{\ \sigma\nu}R^+_{\nu\lambda\sigma\kappa}
-2H_{\kappa}^{\ \sigma\nu}R^+_{\nu\lambda\sigma\beta}$ is
antisymmetric in $\beta$ and $\kappa$. But, after the summation,
these latter indices only can be $u$ and add, therefore, to zero.
Thus,
the generalized Einstein equation
$R^+_{\mu\nu}\=0$ implies hence the vanishing of the
covariant derivatives of any order of
$\nabla^+_{\nu}{R^+}^\nu_{\ \lambda\beta\kappa}\=0$.
Then
${\nabla^+}^\mu{\nabla^+}^\rho R^+_{\mu\nu\rho\sigma}$
vanishes also, since interchanging the order of
the covariant derivatives introduces the
curvature and the torsion, cf. Eq. (2.5), and such terms have already been
shown to vanish.

Let us mention that if $k$ is quadratic in the transverse coordinates,
$k\=k_{ij}(u)x^ix^j$, which corresponds to having
an exact plane wave, the anomaly cancellation can be shown even
non-perturbatively [9], using the standard technique [4-5].

Observe that the vector-potential and the axion enter the formulae (3.3)
with opposite signs, opening the possibility for some cancellations to
take place. Another, even more interesting, solution can be
found in fact by choosing
$$
a_i=\pm b_i,
\equation
$$
so that the $t_{\mu\nu}$ form an upper (lower) triangular matrix. (The
vector-potential can now be an arbitrary function of $x$ and $u$). Choose,
e.g., the plus sign. Then the second eqn. in (3.3) is automatically
satisfied, while the two others reduce to
$$\left\{\eqalign{
R^+_{ju}&=\partial^i\bigg({\partial a_i\over\partial x^j}
-{\partial a_j\over\partial x^i}\bigg)=0,
\ccr
R^+_{uu}&=\partial^i\bigg({\partial a_i\over\partial u}-\2{\partial
k\over\partial x^i}\bigg).
\cr}\right.
\equation
$$
The first of these equations is a {\it vacuum Maxwell equation},
$$
\partial^if_{ij}=0
\where
f_{ij}=\partial_ia_j-\partial_ja_i,
\equation
$$
and the second is
$$
\partial_u{\rm div}\,{\bf a}-\2\Delta k=0.
\equation
$$
Inserting any solution of the vacuum Maxwell equation (3.9) into Eq. (3.10)
we get a Poisson equation for $k$; in the Coulomb
gauge ${\rm div}\,{\bf a}\=0$ one even gets Laplace's equation.
All such solutions  satisfy
the anomaly cancellation condition
at the one-loop level.

Remarkably, our \lq upper-triangular' choice works also at the {\it two-loop}
level. Using the results of Osborn [1], the two-loop contribution to the
anomaly can in fact be worked out to get
$$
-2{R^+}^{\lambda\mu\kappa}_{\ \ \ \ \beta}{R^+}_{\alpha\kappa\lambda\mu}
+3{R^+}^{[\lambda\mu\kappa]}_{\ \ \ \ \ \beta} R^+_{\alpha\kappa\lambda\mu}
+4\big(H^2)^{\kappa\lambda}R^+_{\alpha\kappa\beta\lambda}
+2q\nabla_{\alpha}^+\partial_{\beta}H^2,
\equation
$$
where $q$ is a constant, and
$H^2_{\mu\nu}$, and
$H^2$ was defined earlier. (The term proportional to $q$ can actually
be eliminated
by an
appropriate redefinition of the fields [1]).
From (3.2) one sees that for
the triangular choice $a_i\=b_i$ one has $R^+_{uijk}\=0$, so that the
two first terms in Eq. (3.11) are zero. Next, $\big(H^2\big)^{\mu\nu}$ is
non-zero only for $\mu\=\nu\=v$, $\big(H^2\big)^{vv}$. But $R^+$ with at
least one lower $v$-index is zero, because the metric admits a
covariantly constant null vector. Therefore, the third term vanishes.
Finally, the last term vanishes also, since ${H^2}\=0$ as we have seen
above.

One may wonder if the anomaly vanishes at all
orders without requiring the vector potential to be linear.
This is not completely unlikely, due to
the special structure coming from to the complex structure of Osborn.

One can easily see that
if all higher-order terms of the perturbation expansion of the
$\beta$-function
have (after an appropriate redefinition of the fields) the form
$$
\beta_{\mu\nu}=
{Y^+}^{\lambda\rho\sigma}_{\ \ \ \ \nu}\,{R^+}_{\mu\sigma\lambda\rho}
\equation
$$
for some tensor ${Y^+}^{\lambda\rho\sigma}_{\ \ \ \ \nu}$,
then the triangular configurations are anomaly free at all orders.
The only non-zero components of the Riemann tensor $R^+$ are in fact
$R^+_{ijku}$, $R^+_{iuju}$, and those coming from the
antisymmetry in the first and second pairs of indices.
Thus, one of the summation indices $\sigma\lambda\rho$ should be an $u$;
but then the term in (3.12) is zero as we have seen above.

Another soluble case would arise by replacing Eq. (3.7) by
$a_i\mp b_i\=c_{ij}(u)x^j$ with $c_{ij}$ antisymmetric.

Let us mention finally, that a dilaton can be included as well.

\kikezd{Acknowledgement}. We are indebted to Gary Gibbons and Malcolm
Perry for their interest and collaboration at the early stages of this
work [9], and to J\'anos Balog for discussions.
One of us (Z. H.) would like to thank Tours University for the
hospitality extended to him,
and to the Hungarian National Science and Research Foundation
(Grant No. $2177$) for a partial financial support.

\kikezd{Note added}.
After this work was completed, we discovered a  recent paper by
Tseytlin [10], containing some similar ideas.  
We also learned
that the relation between the axion and
the torsion was considered before by 
E.Braaten,
T.L. Curtright and C.K. Zachos, 
Nucl. Phys. {\bf B260}, 630 (1985),
and that our formula (3.6) 
for the two-loop beta-function appears also in
R.R. Metsaev and A. A. Tseytlin, Nucl. Phys. {\bf B293}, 385 (1987). 
We are indebted to Professor A. Tseytlin for these references.

\vskip4mm


\centerline{\bf\BBig References}

\reference
H. Osborn,
Ann. Phys. {\bf 200}, 1 (1990).

\reference
H. W. Brinkmann,
Proc. Natl. Acad. Sci. U.S. {\bf 9}, 1 (1923);
Math. Ann. {\bf 94}, 119-145 (1925);
see also
C. Duval, G. Gibbons and P. Horv\'athy,
Phys. Rev. {\bf D43}, 3907 (1991).

\reference
D. Kramer, H. Stephani, E. Herlt, M. McCallum,
{\it Exact solutions of Einstein's field equations},
Cambridge : Univ. Press (1980);
J. Ehlers and W. Kundt, {\it Exact solutions of the gravitational field
equations}, in : Witten L. (ed.), Gravitation: an introduction to current
research, Wiley, New York (1962).

\reference
D. Amati and C. Klim\v c\'\i k, Phys. Lett. {\bf B219}, 443 (1989);
R. G\"uven, Phys. Lett. {\bf B191}, 275 (1987);
H. de Vega and N. Sanchez, Nucl. Phys. {\bf B317}, 706 (1989);
R. Rudd, Nucl. Phys. {\bf B352}, 489 (1991);
A. A. Tseytlin, Phys. Lett. {\bf 288}, 279 (1992).

\reference
G. T. Horowitz and A. R. Steif,
Phys. Rev. Lett. {\bf 64} (1990) 260-263; Phys. Rev. {\bf D42}, 1950 (1990);
A. Steif, Phys. Rev.  {\bf D42}, 2150 (1990);
G. T. Horowitz, in Proc. VIth Int. Superstring Workshop
{\it Strings'90}, Texas '90, Singapore: World Scientific (1991).

\reference
M. B. Green, J. H. Schwarz and E. Witten, {\it Superstring Theory}, Vol. 1,
Cambridge University Press, (1987)

\reference
C. Lovelace, Phys. Lett. {\bf B138}, 75 (1884);
E. Fradkin and A. A. Tseytlin,
Phys. Lett. {\bf B158}, 316 (1985); {\bf B160}, 69 (1985);
Nucl. Phys. {\bf B261}, 1 (1985);
C. Callan, D. Friedan, E. Martinec and M. Perry, 
Nucl. Phys. {\bf 262B}, 593 (1985).

\reference
E. Witten, Commun. Math. Phys. {\bf 92}, 451 (1984).

\reference
C. Duval, G. Gibbons, P. A. Horv\'athy, M. J. Perry,
unpublished (1991).

\reference
A. A. Tseytlin,
Nucl. Phys. {\bf B390}, 153 (1993).

\vfill\eject

\bye